\documentclass[superscriptaddress,showpacs,twocolumn,amsmath,amssymb]{revtex4}

\newcommand{\figref}[1]{Fig.~\ref{#1}}

\usepackage{txfonts}
\usepackage{graphicx}
\usepackage{dcolumn}
\usepackage{bm}
\begin{document}
\renewcommand{\baselinestretch}{1}
\title{Excitons in electrostatic traps}
\author{A. T. Hammack}
\affiliation{
Department of Physics, University of California at San Diego, La Jolla, CA 92093-0319
}
\author{N. A. Gippius}
\affiliation{
Department of Physics, University of California at San Diego, La Jolla, CA 92093-0319
}
\affiliation{General Physics Institute RAS, Vavilova 38, Moscow 119991, Russia}
\affiliation{Institute of Physical and Chemical Research (RIKEN), Wako 351-0198, Japan}
\author{G. O. Andreev}
\affiliation{
Department of Physics, University of California at San Diego, La Jolla, CA 92093-0319
}
\author{L. V. Butov}
\affiliation{
Department of Physics, University of California at San Diego, La Jolla, CA 92093-0319
}
\author{M. Hanson}
\affiliation{
Materials Department, University of California at Santa Barbara, Santa Barbara, California 93106-5050
}
\author{A. C. Gossard}
\affiliation{
Materials Department, University of California at Santa Barbara, Santa Barbara, California 93106-5050
}

\begin{abstract}
We consider in-plane electrostatic traps for indirect excitons in coupled quantum wells, where the traps are formed by a laterally modulated gate voltage. An intrinsic obstacle for exciton confinement in electrostatic traps is an in-plane electric field that can lead to exciton dissociation. We propose a design to suppress the in-plane electric field and, at the same time, to effectively confine excitons in the electrostatic traps. We present calculations for various classes of electrostatic traps and experimental proof of principle for trapping of indirect excitons in electrostatic traps.
\end{abstract}



\date{\today}

\maketitle


A possibility of exciton confinement and manipulation in potential traps attracted considerable interest in the past. Excitons are bosonic particles in semiconductors and studies of the bosons in controlled potential reliefs are of fundamental interest. In particular, confinement of bosonic atoms in traps is crucial for experimental implementation of atomic Bose-Einstein condensates \cite{CornellWieman2002, Ketterle2002}. Also, controlling the optical properties of semiconductors by manipulating the excitons in microdevices may be used to develop new optoelectronic devices.

Pioneered by the electron-hole liquid confinement in the strain-induced traps \cite{Wolfe1975}, exciton confinement has been implemented in various traps: strain-induced traps \cite{Trauernicht1983,Kash1988,NegoitaAPL1999}, traps created by laser-induced local interdiffusion \cite{Brunner1992}, magnetic traps \cite{Christianen1998}, and electrostatic traps \cite{Zimmermann1997,Huber1998,Krauss2004}. An advantage of the electrostatic traps is a possibility for creating a variety of in-plane potential reliefs with the required parameters for the excitons and a possibility for manipulating the relief in-situ thus controlling the optical properties both in space and in time. In particular, electrostatically induced transport of excitons \cite{Hagn1995} and the electrostatically controlled capture and release of the photonic images \cite{Krauss2004} have been demonstrated.

The principle of the electrostatic traps is based on the quantum confined Stark effect \cite{Miller1985}: An electric field $F_z$ perpendicular to the QW plane results in the exciton energy shift \mbox{$\delta E = e F_z d$}, where $d$ is the exciton dipole moment. For the indirect excitons in coupled quantum wells (CQWs) \cite{Hagn1995,Huber1998} electrons and holes are separated in different QWs and $d \approx L_{QW}$, the distance between the QW centers. The laterally modulated gate voltage $V_g(x,y)$ creates a laterally modulated electric field and, in turn, a lateral relief of the exciton energy $\delta E(x,y) = e F_z(x,y) d$. Control of $V_g(x,y)$ allows manipulation of the in-plane potential profile for excitons both in space and in time. However, an intrinsic obstacle for exciton confinement in the electrostatic traps is an in-plane electric field $F_r$ that can lead to exciton dissociation \cite{Miller1985}. A strong exciton confinement in the electrostatic traps requires a strong lateral modulation of $V_g$, which, in turn, can lead to a strong $F_r$ and exciton ionization \cite{Zimmermann1997}. In Ref. \cite{Zimmermann1997}, the intensity of the exciton PL was decreased with increasing modulation of $V_g$ indicating that exciton ionization worked effectively against the exciton confinement to the electrostatic traps. 

We modeled electrostatic trap geometries that are variations on the theme of a CQW in an insulating layer sandwiched between a patterned set of top gate electrodes and a single homogeneous bottom gate electrode. To find the field distribution for given voltages at the gate electrodes, the Poisson equation was solved numerically for a static dielectric constant using the boundary elements method \cite{Beer2001}. 

\begin{figure}[t]
\begin{center}
\includegraphics[width=8.6 cm]{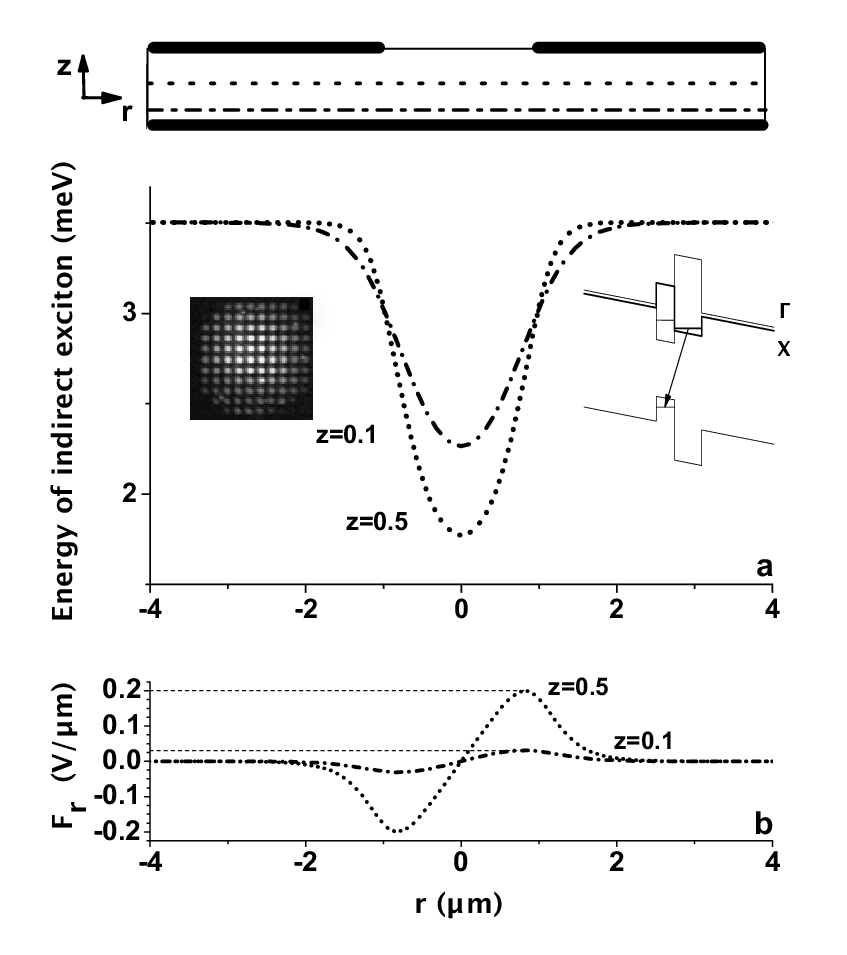}
\caption{
\label{fig:simplewell} 
Schematic cross-section of an electrostatic trap formed by sandwiching a AlAs/GaAs CQW between a homogeneous bottom gate and a top gate with a hole. The indirect exciton effective potential (a) and lateral component of electric field in the CQW plane  
(b) are shown for $V_g=1$ V, $D=1000$ nm, $d=3.5$ nm, and for two CQW positions --- $D_b=$100 nm ($z=D_b/D=0.1$) and 500 nm ($z=0.5$) above the bottom gate. Left inset: the top gate electrode, which forms the periodic trap or bump arrays measured in the experiment. Right inset: AlAs/GaAs CQW band diagram.
}\end{center}
\end{figure}

The first trap is studied for AlAs/GaAs CQWs. It is obtained by making a circular hole in the top gate (\figref{fig:simplewell}). The separation between the gates is taken to be $D$ = 1000 nm and the exciton dipole moment in AlAs/GaAs CQWs is $d \sim 3.5$ nm. \figref{fig:simplewell}a displays the calculated effective exciton potential $\delta E(r) = e F_z(r) d$ for a top gate voltage of 1 V for two positions of the CQW --- 100 nm and 500 nm above the bottom gate. This trap has only two gates and its spacial profile is controlled by the ratio of the hole diameter versus gate separation. The applied voltage modifies only the strength of the trap (at positive $V_g$) or bump  (at negative $V_g$), not its shape. 

The lateral component of the electric field $F_r(r)$ at the CQW plane is shown in \figref{fig:simplewell}b. Positioning the CQW close to the lower gate results in a strong suppression of the in-plane electric field at the CQW plane, by an order of magnitude for the example shown in \figref{fig:simplewell}. This is because the in-plane electric field is concentrated near gate edges and vanishes at the homogeneous bottom gate. The exciton ionization time reduces exponentially with the in-plane electric field $\tau \sim exp[4 E_{ex}/(3 e F_r a_{ex})]$, where $E_{ex}$ is the exciton binding energy and $a_{ex}$ is the exciton Bohr radius (see Eqs. A6 and A7 in \cite{Miller1985} for 2D and 3D excitons). If the CQW is equidistant between the gates, the lateral electric fields reach 0.2 V/$\mu$m (\figref{fig:simplewell}b) and the estimated exciton ionization time (using Eq. A6 in \cite{Miller1985}) is in the ps range and exciton ionization should be prominent. On the other hand, positioning the CQW at the 1/10 distance between the gates, closer to the homogeneous bottom gate, strongly increases the exciton ionization time (by 27 orders of magnitude according to the estimate using Eq. A6 in \cite{Miller1985}) thus making exciton ionization negligible. At the same time $F_z$ is reduced only weakly and the exciton confinement potential remains strong (\figref{fig:simplewell}a).

This design can be employed for creation of trap arrays or bump arrays with the modulation amplitude of the potential energy controlled by a single top gate with a periodic array of holes (\figref{fig:simplewell}). Note that the periodic trap array for excitons, where the modulation amplitude of the potential energy is controlled by a single top electrode, is similar to the optical lattice for atoms, where the modulation amplitude of the potential energy is controlled by the laser intensity \cite{Greiner2002}. The trap or bump arrays for excitons can be employed to study excitons in externally controlled potentials.  Particularly interesting is the possibility to investigate the transition from delocalized to localized excitons (superfluid-insulator transition in the condensate case) with increasing amplitude of the potential, exciton temperature, or density. 

\begin{figure}[t]
\begin{center}
\includegraphics[width=8.6 cm]{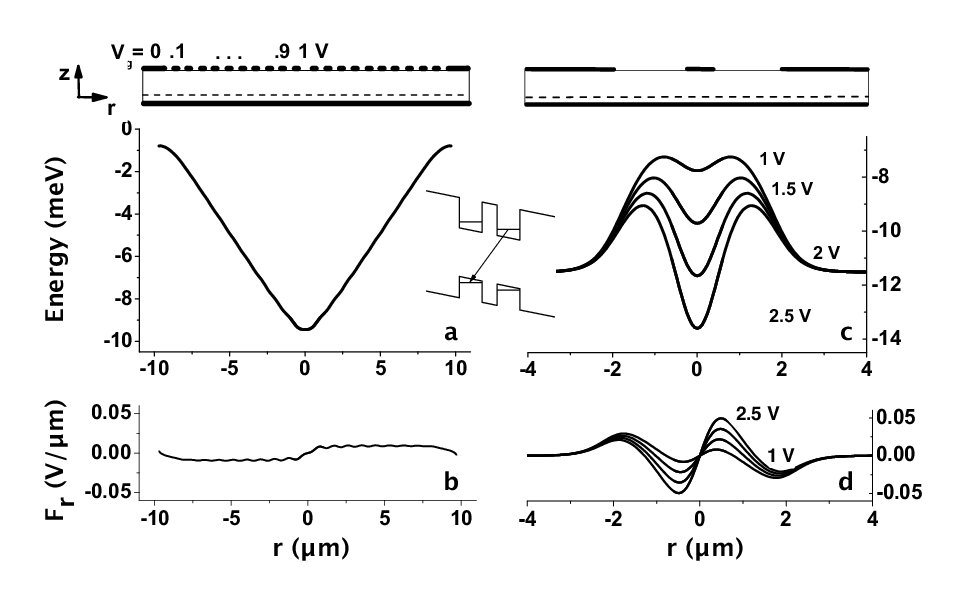}
\caption{
\label{fig:collevap}
The effective potentials and lateral electric fields for a conical trap formed by ten concentric rings with radially decreasing gate voltages $V_g=1 - 0$ V (a,b). A trap for evaporative cooling of excitons for $V_g = 1$ V at external gate and $V_{gc} =$ 1, 1.5, 2, 2.5 V at center gate (c,d). Inset: GaAs/AlGaAs CQW band diagram. $D=1000$ nm, $z=0.1$, and $d=12$ nm. 
}
\end{center}
\end{figure}

The second type of trap is a multi-gated GaAs/AlGaAs CQW structure that allows elaborate control of the radial exciton potential profile (\figref{fig:collevap}a). The top gates comprise a system of ten 200 nm wide concentric rings with a ring width vs inter-ring separation ratio of 2:7. The openings between the gates allow exciton photoexcitation over entire trap area while the opening at the trap center allows optical signal collection for the exciton confined at the trap bottom (this is essential when nontransparent gates are used). An important example of potential trap profiles possible for this geometry is a conical trap for the indirect excitons created by linearly increasing gate voltages toward the center of the structure (\figref{fig:collevap}a). The trap enables collection of a large number of excitons photoexcited over the entire large trap area at the trap bottom. The large exciton number at the trap bottom is essential for studies of exciton BEC in traps since the critical temperature for BEC increases with the exciton density (for review of exciton condensation in confined systems see e.g. \cite{Butov2004}). The gradual reduction of the gate voltage distributed over many gate electrodes allows keeping the in-plane electric field negligibly small (\figref{fig:collevap}b), thus suppressing the exciton ionization.

The third trap type is designed for evaporative cooling of excitons. It has three gates: a bottom gate, an external top gate, and the central top gate (\figref{fig:collevap}c). For central gate voltages $V_{gc} < 1$ V the potential profile is a potential bump for the excitons. However, with the increase of $V_{gc}$ a trap develops at the center of the bump that can be used for evaporative cooling of the indirect excitons: the most energetic excitons overpass the potential barrier and leave the trap thus lowering the temperature of the exciton system in the trap. Evaporative cooling is effectively used for cooling atomic gases in traps \cite{CornellWieman2002, Ketterle2002}. 

Following the computational modeling of the electrostatic exciton traps we fabricated samples based on the first design type (\figref{fig:simplewell}). The studied electric field tunable $n-i$ AlAs/GaAs CQW structure was grown by MBE. The bottom $n^+$ layer is Si-doped GaAs with $N_{Si} = 2\times10^{18}$ cm$^{-2}$. It serves as a homogeneous bottom gate. Unlike samples studied earlier \cite{Butov2004}, a surface $n^+$ layer was not grown for this sample to allow patterning the top gate by evaporated nontransparent metal contacts. The $i$-region consists of a 2.5 nm GaAs layer and a 4 nm AlAs layer surrounded by two Al$_{0.48}$Ga$_{0.52}$As barrier layers with thicknesses of 900 nm and 100 nm for the upper and lower barriers. Mesas were etched and a patterned top metal electrode containing an array of circular holes with diameters 10 $\mu$m and periods 20 $\mu$m was deposited onto the mesas. A broad 200 $\mu$m open area in the top electrode was fabricated as well for comparison. The sample was excited by cw 532 nm laser and the PL measurements were performed at $T=1.6$ K.

First we note that essentially no change of the exciton energy in the open area vs gate voltage was detected. This indicates that consistent with our calculations an electric field is present in the CQW only in regions close to the upper gate.  

\begin{figure}[t]
\begin{center}
\includegraphics[width=8.6 cm]{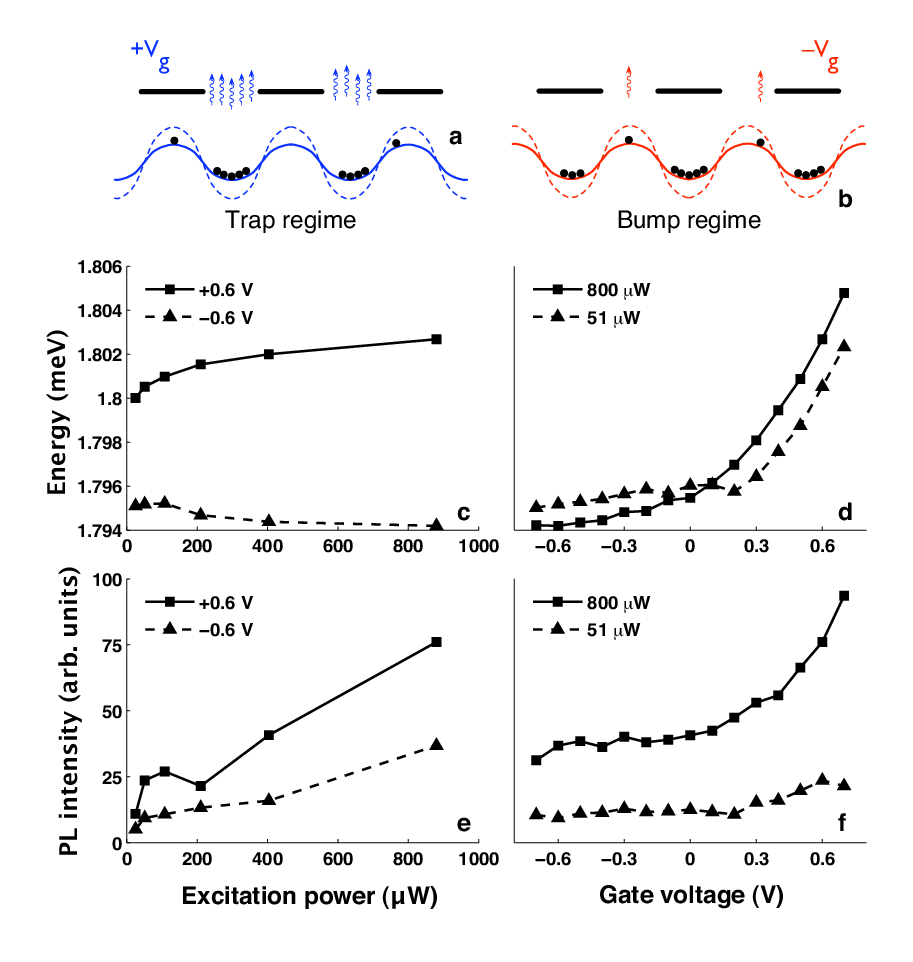}
\caption{\label{fig:exp}
Experimental proof of principle for electrostatic trapping of excitons. 
Top panel: scheme of the potential profiles for positive $V_g$, the trap regime (a), and negative $V_g$, the bump regime (b). The dependence of the PL line energy and intensity on laser excitation power at $V_g=0.6$ and -0.6 V (c,e). The dependence of the PL line energy and intensity on gate voltage for two excitation powers (d,f).
}
\end{center}
\end{figure}

The results for the indirect exciton PL for the array of 10 $\mu$m holes as a function of gate voltage and excitation power are presented in \figref{fig:exp}.  For the CQW type and contact geometry studied, positive applied top gate voltages should cause formation of potential traps beneath the holes in the gate (\figref{fig:exp}a), while negative applied voltages should lead to potential bumps beneath the holes (\figref{fig:exp}b). \figref{fig:exp}c shows that increasing the exciton density leads to enhancement of the PL energy in the trap regime and its reduction in the bump regime. This corresponds to the expected behavior: The indirect excitons are oriented dipoles and interaction between them is repulsive \cite{Butov2004}. The repulsive interaction leads to enhancement of the exciton energy in the area where the excitons accumulate and, in turn, to reduction of the exciton energy in between the areas of exciton accumulation \cite{Ivanov2002} --- the repulsively interacting excitons screen the external potential. In the trap regime, the excitons accumulate in the potential traps beneath the holes in the top gate (\figref{fig:exp}a) and the increasing exciton density in the traps is observed in the PL energy enhancement (\figref{fig:exp}c). In contrast, in the bump regime, the excitons accumulate beneath the areas covered by the gates (\figref{fig:exp}b) and the increasing exciton density outside the holes in the top gate is observed in the PL energy reduction (\figref{fig:exp}c). The exciton accumulation in the traps beneath the holes in the top gate is revealed also by the stronger intensity enhancement in the trap regime compared to that in the bump regime (\figref{fig:exp}e). 

The transition from the enhancement of the PL energy with density to its reduction is observed around $V_g = 0$ V (\figref{fig:exp}d). This is expected since the transition from the trap regime to the bump regime takes place at $V_g = 0$ V for the AlAs/GaAs CQWs. Finally, the formation of the electrostatic exciton traps by the laterally modulated gate voltage at positive $V_g$ is accompanied by enhancement of the exciton PL energy and intensity (\figref{fig:exp}d, f) that is typical for the exciton accumulation in the traps. The observed enhancement of the exciton PL intensity and energy with the electrostatic trap formation shows that the exciton ionization due to the in-plane electric field is suppressed and the excitons accumulate in the traps.

In conclusion, we proposed a design to suppress the in-plane electric field and at the same time to effectively confine excitons in the electrostatic traps. This is achieved by positioning the QW plane closer to the laterally homogeneous gate, for instance at the 1/10 distance between the homogeneous bottom gate and patterned top gate. Then, we presented calculations for various types of the electrostatic traps: traps for building the trap or bump arrays, traps for accumulation of a large exciton number, and traps allowing evaporative cooling for excitons. Finally, we presented experimental proof of principle for trapping of indirect excitons in electrostatic traps.

The authors are thankful to D.S. Chemla, T. Ishihara, L.S. Levitov, and S.G. Tikhodeev for useful discussions. N.A.G. thanks Russian Foundation for Basic Research and Russian Ministry of Science for partial financial support.

\end{document}